QUANTUM FLUCTUATIONS CONTRIBUTION TO THE RANDOM WALK OF A SINGLE MOLECULE AND NEW ESTIMATE OF THE PLANCK CONSTANT

Jean Paul Mbelek[1]

1. Service d'Astrophysique, CEA-Saclay, Orme des Merisiers, 91191 Gif-sur-Yvette, France

Correspondence to: Jean Paul Mbelek[1] Correspondence and requests for materials should be addressed to J.P.M. (Email: mbelek.jp@wanadoo.fr).

Abstract : It is shown, by considering the case of the harmonic oscillator, that quantum fluctuations may be the most significant contribution to the random walk of a single molecule. From this point, the controversy on the existence of a "standard quantum limit" (SQL) is addressed and settled on the experimental ground. Comparisons to the experimental data yet avalaible in the literature provide a new estimate of the reduced Planck constant yielding $\hbar = (1.1 \pm 0.2) \times 10^{-34}$ J.s.

I. Introduction

The uncertainty principle was first introduced in quantum mechanics by W. Heisenberg[1] to express quantitatively the necessity to give up classical notions like the orbits of electrons in atoms. A more rigorous and general mathematical derivation of the uncertainty relations was next provided by E. H. Kennard[2] then H. P. Robertson[3], by defining properly the uncertainties $\Delta x$ and $\Delta P_x$ as standard deviations. These authors show in particular that the commutation relation $[x , P_x] = i \hbar$ implies $\Delta x \Delta P_x \geq \frac{1}{2} |<[x , P_x]>|$ or otherwise stated $\Delta x \Delta P_x \geq \hbar/2$, the equality $\Delta x \Delta P_x = \hbar/2$ being obtained for the harmonic oscillator. However, the definition of the standard deviation may change from one experimental situation to another. So, carrying the analysis on the uncertainty principle, D. Bohm[4] gave an estimate of the time variation of the fluctuation $\Delta x$ of the position x of a particle (§ 5.4) which reads

$$\Delta x = \hbar\, t/m\, \Delta x_0 \qquad (1)$$

and also follows from the spread of the wave packet of initial width $\Delta x_0$ in the limit $t \to \infty$ (see § 3.13). Now, by assuming $|\Delta x - \Delta x_0| << \Delta x_0/2$, clearly relation (1) yields $\Delta x^2 \cong (\hbar/m)\, t$. Further, the complete expression $\Delta x = \Delta x_0 [1 + (\hbar t/m\Delta x_0^2)^2]^{1/2}$ that holds at any time implies $\Delta x \geq \Delta x_0$ and $\Delta x^2 \geq (\hbar/m)\, t$, therby suggesting a SQL. However, let us notice that by setting

$$\Delta(dx/dt) = d\Delta x/dt \qquad (2)$$

and considering the case of the harmonic oscillator, it is straightforeward to derive after integration the equality

$$\Delta x^2 = (\hbar/m)\, t \qquad (3)$$

On the other hand, still considering the case of the harmonic oscillator but assuming $d\Delta x/dt \leq \Delta P_x/m$, one would obtain instead
$$\Delta x^2 \leq (\hbar/m)\, t \qquad (4)$$

without necessarily being in conflict with the uncertainty relations. Relation (3), where t stands for the averaging time, has been put foreward (up to a factor 2 in ref. 6) by V. B. Braginsky et al.[5,6,7] in their study of the possibility to evade the confrontation with the uncertainty relations. Nevertheless, H. P. Yuen[8] pointed out that the derivation of relation (3) from the uncertainty principle is incorrect, and even needs not hold at all. Relation (4) seems rather consistent with the claims by H. P. Yuen[9, 10], M. Ozawa[11, 12] and M.T. Jaekel and S. Reynaud[13]. Moreover, we have found some experimental cases in the literature[14,15,16,17,18] which plead for relation (4). It turns out that the controversy on the existence of a SQL[9, 10, 11, 12, 13,19], which is of importance for the interferometric detection of gravitational waves, may be settled on the experimental ground. It already seems that the SQL does not hold. As stated above, either relation (3) or (4) may apply, depending on how $d\Delta x/dt$ compares to $\Delta v_x = \Delta P_x/m$. In the following, we shall consider the case of the harmonic oscillator where the avoidance of the SQL is realized. Thence, by comparing the theoretical results obtained in this study with the avalaible experimental data, an estimate of the reduced Planck constant is derived.

II. Quantum fluctuations contribution to the random walk

As one can see, relation (3) compares to the 1D Brownian motion relation $\Delta x^2 = 2D_0\, t$. Hence, assuming that the condition (2) is realized, by taking into account both the classical and quantum contribution to the random motion of a particle, we predict a total diffusion coefficient

$$D = D_0 + \hbar/2m \qquad (5)$$

Let us emphasize that the quantum contribution $\hbar/2m$ to the diffusion coefficient is supported by the Schrödinger equation - $(\hbar^2/2m)\, \Delta\psi = i\hbar\, \partial\psi/\partial t$ of a free particle of mass m. Indeed, the latter rewrites in imaginary time $\tau = i\, t$ : $(\hbar/2m)\, \Delta\psi = \partial\psi/\partial\tau$, which looks like the diffusion equation with a diffusion coefficient equal to $\hbar/2m$.

III. Comparison with experimental data and estimate of the reduced Planck constant

Hereafter, we consider various experimental situations where a single molecule is both subject to harmonic oscillations and diffusion along a polymer chain. While the molecule under consideration oscillates and undergoes a random walk, it sweeps out a volume V defined by a lateral surface S and the displacement x. Since quantum fluctuations occur, those of the latter geometrical quantities amount respectively to $\Delta V = \Delta x\, \Delta y\, \Delta z$, $\Delta S = \Delta y\, \Delta z$ and $\Delta x$. Now, one may write accordingly

$$d\Delta V = \Delta S \times \Delta v_x \times dt \qquad (6)$$

Hence, the 1D mean-square displacement (MSD) reads finaly

$$\Delta z^2 = \Delta y^2 = \Delta x^2 = (\hbar/3m)\, t \qquad (7)$$

which is less then the SQL by a factor 3. So, the total diffusion coefficient rewrites

$$D = D_0 + \hbar/6m \quad (8)$$

Table 1 and Figure 1 below give respectively some experimental data found in the literature and the fit to these data. The results are found in good agreement with relations (7) and (8) by neglecting the classical diffusion coefficient, $D_0$, with respect to the quantum mechanical contribution $\hbar/6m$. Also, we have found one case in the literature[22] where the lower bound of D is equal to $\hbar/6m$, the upper bound of $D_0$ is then estimated to 5.22 µm²/s on account of relation (8). Let us emphasize that diffusion coefficients that are less than $\hbar/2m$, in consistency with relation (4) by neglecting $D_0$, have been reported too in the literature[14]. All other data, we have found in the literature[20,21] are consistent with relations (7) and (8) by neglecting instead $\hbar/6m$ with respect to $D_0$.

| M (kDa) [Ref.] | D (µm²/s) [Ref.] |
|---|---|
| 1050 [15] | 0.0082 ± 0.0014* [15] |
| 113 [16] | 0.11 ± 0.01 [16] |
| 61 [16] | 0.18 ± 0.02 [16] |
| 61 [17] | 0.16 ± 0.03* [17] |
| 50 [18] | 0.23 ± 0.08 [18] |

Table 1 : 1D Diffusion coefficient, D, of the Brownian motion of a single molecule of mass m. The data are taken from the references. The uncertainties* of the first and fourth raws have been derived by us from the error bars of the experimental curves of the authors (1 Da = 1,66054 × 10⁻²⁷ kg ; 1 kDa = 10³ Da).

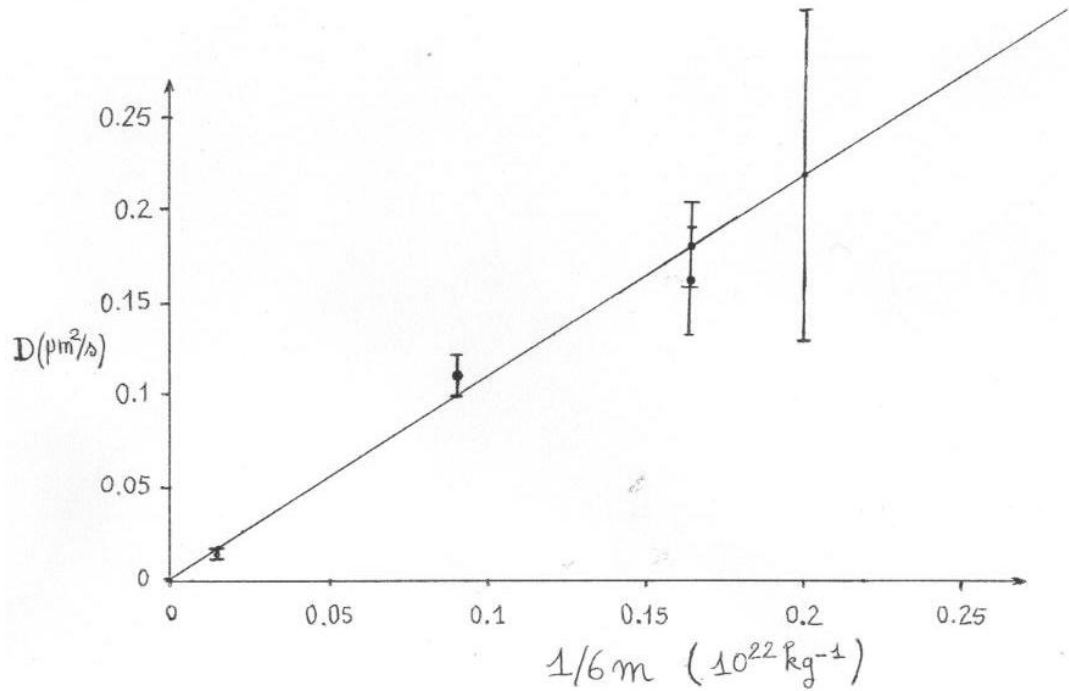

Figure 1 : 1D diffusion coefficient D (µm²/s) versus 1/6m (10²² kg⁻¹). The linear fit to the data (see table 1) yields a slope equal to (1.1 ± 0.2) × 10⁻³⁴ J.s (reduced Chi-square $\chi_v^2$ = 1.19), which is a fairly good estimate of $\hbar$ as compared to the 2006 CODATA value $\hbar$ = (1.054 571 628 ± 0.000 000 053) × 10⁻³⁴ J.s.

IV. Conclusion

We have shown that quantum fluctuations contribute to the random walk of a single molecule. In addition, referring to the experimental data, it turns out that the SQL is evaded in most cases involving the random walk of a single molecule. As a consequence, we have derived a fairly good estimate of the reduced Planck constant equal to $\hbar = (1.1 \pm 0.2) \times 10^{-34}$ J.s.